\documentclass[prl,twocolumn,showpacs,superscriptaddress,floatfix]{revtex4}
\usepackage{graphicx}

\newcommand{\be}{\begin{equation}}
\newcommand{\ee}{\end{equation}}
\newcommand{\bea}{\begin{eqnarray}}
\newcommand{\eea}{\end{eqnarray}}
\newcommand{\down}{\downarrow}
\newcommand{\up}{\uparrow}
\newcommand{\f}{\frac}

\begin{document}

\title{Length-Dependent Conductance of a Spin-Incoherent Hubbard Chain: Monte Carlo Calculations}

\author{Olav  F. Sylju{\aa}sen}

\affiliation{NORDITA, Blegdamsvej 17, DK-2100 Copenhagen {\O}, Denmark}


\preprint{NORDITA-2006-33}

\begin{abstract}
The dc conductance of a short spin-incoherent Hubbard chain coupled to leads is investigated using quantum Monte Carlo calculations. In contrast with the Luttinger liquid regime, where the conductance is equal to the non-interacting value,  the spin-incoherent regime displays a conductance that decreases rapidly with chain length down to a value of roughly $1.5e^2/h$ for a four site chain followed by a slower decrease for longer chains. We also discuss the resistance contribution from scattering in the contacts. 
\end{abstract}

\maketitle

Advances in making small electronic devices have made it increasingly important to understand the quantum mechanics of confined electrons. Especially important is the quantum wire where the electrons are restricted to move in a narrow channel. When such a quantum wire is made thinner its conductance decreases in perfect steps of $2e^2/h$\cite{CondQuant}. There is, however, an anomaly in this otherwise perfect picture; at $0.7 \times 2 e^2/h$ an extra steplike structure is seen at very low electron densities at finite temperatures\cite{07}. While the regular steps can be explained in terms of 
non-interacting electrons, different theories have been proposed for the anomalous step\cite{Matveev,Meir,Berggren,Reilly}.  
A particularly intriguing theory\cite{Matveev} is that the suppressed conductance is a result of the electron gas entering a {\em spin-incoherent} (SI) regime\cite{VadimMisha1} at high enough temperatures. It is however unclear how the prefactor $0.7$ comes about within this theory. Experiments indicate that this prefactor depends on the length of the wire\cite{Length}. However, not much is known about this theoretically. The aim of this Letter is to investigate this length dependence numerically.

At zero temperature, $T=0$, the low-energy model of a generic interacting 1D electron liquid is the Luttinger liquid (LL) which is characterized in part by having independent spin and charge excitations each having separate coupling constants, characteristic velocities and bandwidths. 
The conductance of a uniform pure LL depends on the charge sector coupling constant $K_c$ alone and is at low temperatures $G=2K_c e^2/h$\cite{ApelRice,KaneFisher}. However this result does not coincide with what is being measured in experiments where leads are inevitably connected to the interacting electron liquid. The leads have a profound influence on the dc conductance and it has been shown by modeling the leads as 1D non-interacting electron gases that the conductance remains at the non-interacting value $2e^2/h$ provided one couples the leads sufficiently smoothly to the interacting wire\cite{nonrenormalized}. Otherwise the conductance is solely determined by scattering in the contacts. As will be shown the presence of leads also influences the conductance in the SI regime.

For temperatures higher than the bandwidth of either the spin or the charge excitations one do not expect the LL description to hold. In general the spin bandwidth $J$ is smaller than the charge bandwidth $E_f$ when the interaction energy dominates. Thus a situation where $J \ll T \ll E_f$ is conceivable. This is known as the SI regime, where in contrast to the LL regime, the electron Green function displays non-propagating spin excitations\cite{VadimMisha1,Fiete}, a broad momentum distribution\cite{VadimMisha3}, and an anomalous density of states\cite{Tunneling,ZeroBias}.  The boundary Green function has also been explored\cite{Kakashvili} as well as Coulomb drag effects\cite{CoulombDrag}, the Fermi edge singularity\cite{FermiEdge}  and transport properties in the presence of impurities\cite{Transport}. It was argued in Ref.~\cite{Matveev} that spin-charge separation gets violated when leads are coupled to a SI wire, and as a consequence the dc conductance is renormalized to a value $e^2/h$\cite{Matveev}. While the result in Ref.~\cite{Matveev} was obtained for an electron liquid at low densities forming a Wigner crystal, it has been argued that the physics of the SI regime is largely independent of microscopic details, and that one might as well replace the electron gas with a model having short-range interactions like the  Hubbard model. This was utilized in Ref.~\cite{KindermannBrouwer} where the non-equilibrium conductance and noise of a SI chain coupled to leads was calculated in two regimes, a strongly biased regime and a regime with dissipative spin damping, both giving a conductance $e^2/h$. 

\begin{figure}
\includegraphics[clip,width=8cm]{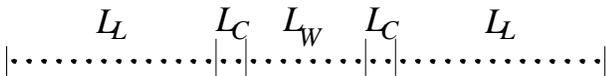} 
\caption{The 1D lattice with noninteracting lead regions of length $L_L$, contact regions of length $L_C$, and the fully interacting chain of length $L_W$.  
\label{chainfig}}
\end{figure}

We model the SI wire coupled to leads as a Hubbard model with a site dependent potential $u_i$
\be \label{Eq:Hubbard}
H = -t  \! \! \! \! \! \! \sum_{\langle ij \rangle,\sigma=\{\up,\down\}} \! \! \! \! \! \left( c_{i \sigma}^\dagger c_{j \sigma} + c_{j \sigma}^\dagger c_{i \sigma} \right) + \sum_i u_i n_{i\up} n_{i\down},
\ee
where $c_{i \sigma}$ is the spin $\sigma$ fermion annihilation operator at site $i$, and $n_{i\sigma}$ is the corresponding density operator. 
We treat a 1D lattice with $L$ sites using open boundary conditions and divide it into five regions, see Fig.~\ref{chainfig}: Two non-interacting lead regions each of length $L_L$ where $u=0$, an interacting chain region where $u=U$ and two contact regions each of length $L_C$ where $u$ is site-dependent so that the particle density in the chain region interpolates roughly linearly to the density in the lead. Throughout this Letter we add a uniform chemical potential $\mu=-0.3t$ causing the density in the lead regions to be $\sim 0.9$, slightly less than half-filling. The temperature is $T = t/80$ and $L=192$.  To avoid complications due to Kondo physics, we keep $L_W$ even, yet we expect that our results also apply to odd $L_W$ above the Kondo temperature. 

For the Hubbard model the bandwidth of the spin excitation spectrum is 
$J=(4t^2/U) n [1-\sin( 2\pi n)/2\pi n]$, 
while the bandwidth of the charge spectrum is proportional to the Fermi energy $E_f$. The SI regime $J \ll E_f$ is thus realized for low densities, or equivalently for large $U$. In addition $T$ must be placed in-between these scales. The latter is more demanding for low densities than for large $U$, because of the smallness of $E_f$ at low densities, thus requiring very low temperatures which are costly using the numerical technique employed here.  Therefore we will model a chain in the SI regime by setting $U=\infty$ in the interacting region. This implies $J=0$ even for high densities, thus $E_f$ can be kept of the order $t$.  
 
In order to investigate the 1D inhomogeneous system Eq.~(\ref{Eq:Hubbard}) we employ the Stochastic Series Expansion (SSE) quantum Monte Carlo (QMC) method with directed-loop updates\cite{SS}. Fermion QMC simulations frequently come with a minus-sign problem. However this is avoided in 1D using open boundary conditions. The fermions are represented in the occupation number formalism following the sign convention described in Ref.~\cite{Sandvik}, and the directed-loop rules were taken from Ref.~\cite{Directedloops}. In addition to having updates that add/remove a particle with a certain spin we also employ moves that flips the spin of a particle. This is necessary to ensure short autocorrelation times in the SI regime.

The dc conductance $G$ is evaluated as the linear response of the current operator $j_x=i(et/\hbar) \sum_{\sigma=\up,\down}(c^\dagger_{x\sigma} c_{x+1\sigma}-c^\dagger_{x+1\sigma} c_{x\sigma})$ at position $x$ to a discontinuity in the chemical potential at $y$ in the limit of zero frequency
\be
    G =  \lim_{z \to 0} g(z), \; \; \; \;  
    g(z) = Re \f{i}{\hbar} \int_0^\infty dt e^{izt} \langle \left[ j_x(t),P_y \right] \rangle 
\ee
where $P_y$ is the sum of fermion charge density operators at sites to the right of $y$, $P_y=e \sum_{y^\prime > y} n_{y\prime}$. The extrapolation to zero frequency can be taken along any path in the complex plane. We will extrapolate along the imaginary axis\cite{imaginary}, thus $z$ is taken to be imaginary and is denoted $z=i\omega$. Using the charge/current continuity relation for a 1D system with open boundary conditions one finds that $g(i\omega)$ can be evaluated at the Matsubara frequencies $\omega=\omega_n \equiv 2\pi n T$ as\cite{LouisGros}
\be
    g(i\omega_n) = \f{\omega_n}{\hbar} \int_0^{\beta \hbar} d\tau \cos (\omega_n \tau) \langle P_x(\tau) P_y(0) \rangle.
\ee
This correlation function which involves only density operators is easily evaluated using the QMC method. While for a finite system, $g(i\omega=0)=0$, the correct way of obtaining the conductance in the thermodynamic limit is to first take the infinite system size limit and then extrapolate to zero frequency. For a big enough system the finite Matsubara frequencies is not appreciably affected by the system size, thus the dc conductance in the thermodynamic limit can be gotten from extrapolating the conductance at the finite Matsubara frequencies to zero along the imaginary frequency axis. 
 \begin{figure}
\includegraphics[clip,width=8cm]{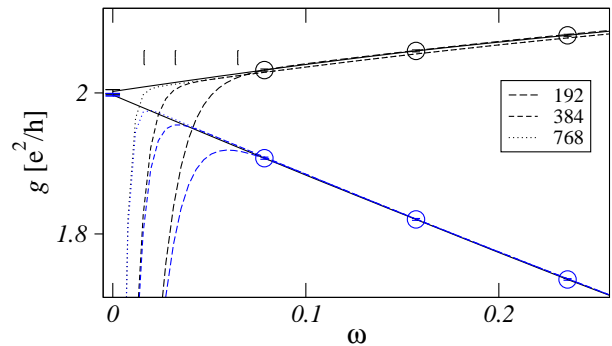} 
\caption{(Color online) Imaginary frequency conductance for free fermions. The dotted, dashed and long dashed lines are exact results for different system sizes $L$ indicated by the legends. The circles are QMC data at the three lowest non-zero Matsubara frequencies for $L=192$. The solid lines are the extrapolated QMC results using the prescription in the text and the error bars at $\omega=0$ indicate the dc conductance in the thermodynamic limit. The upper set of curves (black) is for $x=y=L/2-1$ and the lower (blue) is for $x=L/2-1$, $y=x+1$. The three vertical bars indicate the frequency $\omega=2\pi \hbar v_f/L$ for the three system sizes $L=768,384,192$. 
\label{freebeta80}}
\end{figure}
Figure \ref{freebeta80} shows results for free fermions. The two set of curves are for different choices of $x$ and $y$. Both lead to a value consistent with the exact value $2e^2/h$. The circles are QMC data and the extrapolations are gotten by constructing a rational polynomial function of degree $[p/q]$ that coincides with the QMC data at the lowest $p+q+1$ Matsubara frequencies. As the QMC data has error bars this construction is bootstrapped over $10^4$ repetitions, and the median value is recorded. The median is chosen instead of the average to minimize the effects of spurious poles. This is repeated for all $[p/q]$ with $5 \leq p+q \leq 8$ and $p,q \geq 2$. The solid line shows the average of these (median) values and the error bar at $\omega=0$ shows the corresponding maximum spread, which is larger than the error of a single $[p/q]$ extrapolation.

 The dashed and dotted curves that approach 0 for low $\omega$ are exact results on chains of different lengths. They start to deviate from the extrapolated results at $\omega \approx \hbar v_f 2\pi/L$ which correspond to frequencies lower than the level spacing of the system. This sets a lower limit, $L > \hbar v_f \beta  $, for the system size needed at a finite temperature $T=1/\beta$, in order to ensure that the finite Matsubara frequencies all reflect the infinite size limit. For $T=t/80$ a system size of $L=192$ is sufficient. 

\begin{figure}
\includegraphics[clip,width=8cm]{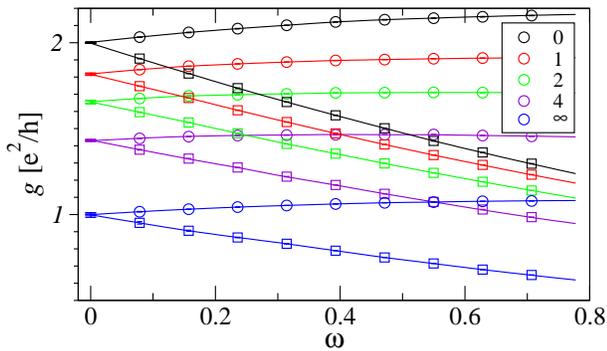}
\caption{(Color online) Imaginary frequency conductances extrapolated to zero frequency for uniform Hubbard chains with different values of $U$. From top to bottom $U/t=0,1,2,4,\infty$. $x=L/2-1$. The circles are for $y=x$ and the squares are for $y=x+1$.
\label{uniform}}
\end{figure}

To check our method we obtain the dc conductances for uniform chains with no leads having different values of the interaction $U$. The measured particle densities are $0.903,0.781,0.696,0.601,0.457$ for $U/t=0,1,2,4,\infty$ respectively. Fig.~\ref{uniform} shows the results containing two different extrapolations for each value of $U$. Taking the average of the extrapolated results we obtain $G/(e^2/h)=1.999 \pm 0.004, 1.817 \pm 0.008, 1.656 \pm 0.012, 1.431 \pm 0.006, 0.999 \pm 0.014$ for $U/t=0,1,2,4,\infty$ respectively. These values agree well with $G=2K_c e^2/h$, where $K_c$ is the Luttinger liquid charge sector coupling constant whose value can be obtained for the Hubbard model for a given density and $U$ by the Bethe Ansatz\cite{Schulz}.   

\begin{figure}
\includegraphics[clip,width=8cm]{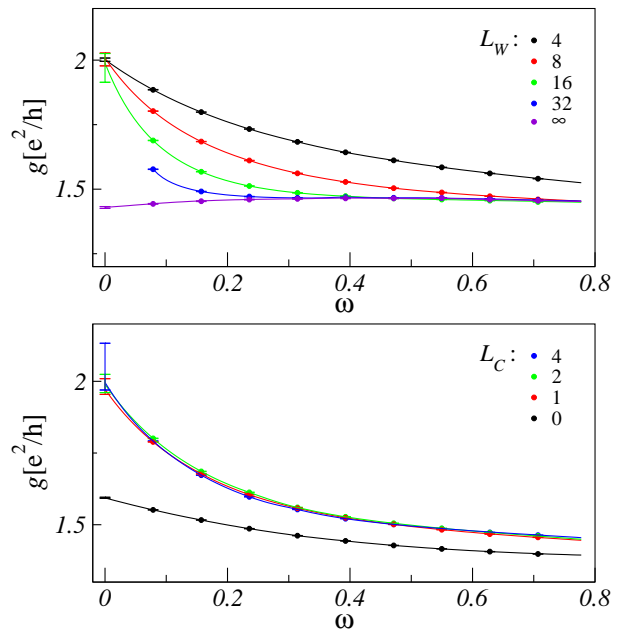}
\caption{(Color online) Imaginary frequency conductances for interacting $(U=4t)$ chains with leads. Upper panel: The curves are for different $L_W$ indicated by the legends and a fixed $L_C=2$. For comparison the uniform $U=4t$ chain without leads is also shown (labeled $\infty$). Lower panel: The curves are for different $L_C$ indicated by the legends and a fixed $L_W=8$.
In both panels $x=y=L/2-1$.
\label{luttleads}}
\end{figure}

We now attach leads to the $U=4$ Hubbard chain. Figure \ref{luttleads} upper panel shows the imaginary frequency conductances and their extrapolations to zero frequency for different lengths $L_W$ of the interacting region. In order to make the contact resistance small we have made contact regions of length $L_C=2$.
While at high Matsubara frequencies the conductance curves are close to the conductance curve of the uniform chain without leads, there is a change in behavior when the frequency gets below $\omega \sim \hbar v_f/ L_W$, where the conductance curves extrapolate towards the non-interacting value in accordance with Refs.~\cite{nonrenormalized}.
To see how the value of the dc conductance depends on the length of the contact regions we show in Fig.~\ref{luttleads} lower panel the conductance for different lengths of the contact regions keeping the length of the interacting chain fixed. The curve with an abrupt contact region ($L_C=0$) shows a lower dc conductance than the others which appear to have reached the adiabatic limit already for $L_C=1$. 

\begin{figure}
\includegraphics[clip,width=8cm]{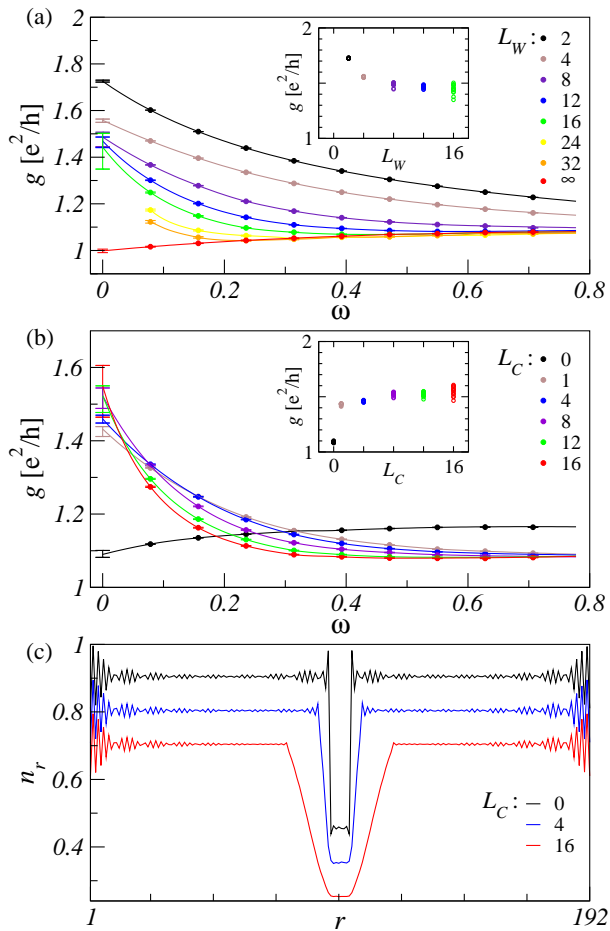}
\caption{(Color online) Imaginary frequency conductances of SI chains with leads. $y=x=L/2-1$, $L=192$. (a) curves are for different $L_W$ indicated by the legends, $L_C=2$. The inset shows the dc conductances. (b) Fixed $L_W=8$ and different $L_C$ indicated by legends. Inset: dc conductances. (c) particle density vs. site for different $L_C$. $L_W=8$. The $L_C=4$$(16)$ curve has been shifted downwards by 0.1(0.2) for clarity.   
\label{uinf}} 
\end{figure}

The SI limit, $U=\infty$, is reached by restricting the Hilbert space so that no sites in the interacting region are doubly occupied. 
Fig.~\ref{uinf}(a) shows the imaginary frequency conductances and their extrapolations for SI chains of different lengths $L_W$ coupled to leads. As in the LL regime, the data at low imaginary frequency differ from that of the uniform chain without leads (labeled $\infty$) and extrapolate to values larger than $e^2/h$. The extrapolation for the longest chains are omitted because only the few lowest Matsubara frequencies distinguish these curves from the uniform one causing excessively large error bars. Nevertheless it is quite clear that the dc conductances of the longest chains also extrapolate to values larger than $e^2/h$. This is rather remarkable in view of the exponential decrease of the single-particle Green function in the SI regime.  The extrapolated dc conductances are plotted in the inset, one datapoint for each $[p/q]$, and reveal a rapid decrease with increasing chain length to a value $\sim 1.5 e^2/h$ for $L_W=4$, and then a further slower decrease.  Note that this intermediate conductance value is close to the value relevant to explain the $0.7$ conduction anomaly.  

In addition to the intrinsic resistance of the interacting region, our results include possible contributions from scattering in the contact regions. The contact regions used were of length $L_C=2$, which is long enough to give adiabatic contacts in the LL regime, see Fig.~\ref{luttleads}. To check their role in the SI regime we show in Fig.~\ref{uinf}(b) results for different $L_C$. While there is a significant resistance contribution for abrupt junctions, $L_C=0$, there are only small differences between the dc conductances for contacts of lengths $L_C \ge 1$, see inset. It is not entirely clear however, due to the large error bars for the largest values of $L_C$, if the dc conductance has saturated or will keep increasing for even larger $L_C$'s. Assuming saturation we estimate that the contact contribution to the reduced conductance is $\sim 0.1 e^2/h$ for $L_C=2$. The density variations in the contacts causing scattering is shown in Fig.~\ref{uinf}(c).  Note the rather abrupt density changes that cause resistance in the case $L_C=0$, compared to the smoother density variations for $L_C=16$. 

Our results show that the dc conductance decreases rapidly with $L_W$ to roughly $1.5 e^2/h$ for $L_W=4$ and then decreases further slowly with $L_W$ for $L_W >4$. It is not clear what sets the rather long length scale associated with this slow decrease. It is plausible that this is related to the magnitude of terms in the effective Hamiltonian breaking spin-charge separation. While it is intriguing that the conductance values obtained here are consistent with the value $0.7 \times 2$ for a broad range of wire lengths $L_W$, it remains to be seen to which extent these values are universal. 

The simulations were in part carried out using computers provided by the University of Aalborg, Denmark and the University of Tartu, Estonia using the NorduGrid ARC middleware.

\end{document}